\documentclass[
aps,%
11pt,%
final,%
notitlepage,%
oneside,%
onecolumn,%
nobibnotes,%
nofootinbib,%
superscriptaddress,%
noshowpacs,%
centertags]%
{revtex4}

\def\squareforqed{\hbox{\rlap{$\sqcap$}$\sqcup$}}

\def\sq{\ifmmode\squareforqed\else{\unskip\nobreak\hfil
\penalty50\hskip1em\null\nobreak\hfil\squareforqed
\parfillskip=0pt\finalhyphendemerits=0\endgraf}\fi}

\def\utw{\smash{\rlap{\lower5pt\hbox{$\sim$}}}}

\def\udtw{\smash{\rlap{\lower6pt\hbox{$\approx$}}}}

\def\diameter{{\ifmmode\mathchoice
{\ooalign{\hfil\hbox{$\displaystyle/$}\hfil\crcr
{\hbox{$\displaystyle\mathchar"20D$}}}}
{\ooalign{\hfil\hbox{$\textstyle/$}\hfil\crcr
{\hbox{$\textstyle\mathchar"20D$}}}}
{\ooalign{\hfil\hbox{$\scriptstyle/$}\hfil\crcr
{\hbox{$\scriptstyle\mathchar"20D$}}}}
{\ooalign{\hfil\hbox{$\scriptscriptstyle/$}\hfil\crcr
{\hbox{$\scriptscriptstyle\mathchar"20D$}}}}
\else{\ooalign{\hfil/\hfil\crcr\mathhexbox20D}}%
\fi}}

\begin{document}
\selectlanguage{english}

     \title{ Dwarf Spheroidal Galaxies in the M101 group and behind it}

                \author{\firstname{I.~D.} ~\surname{Karachentsev}} 
                \author{\firstname{L.~N.} ~\surname{Makarova}}

 \affiliation{ Special Astrophysical Observatory, Russian Academy of Sciences, Russia}
 
 \begin{abstract}
  We use images from the Hubble Space Telescope to determine the distances
to a dozen objects with low surface brightness recently observed around the
bright nearby spiral M101. Only two dwarf galaxies, M101-DwA and M101-Dw9,
turn out to be actual satellites of M101 at distances of about 7 Mpc. The
other objects are probably members of a distant group around the S0-galaxies
NGC~5485/5473. Based on the radial velocities and projected separations of
the 9 satellites, we obtain an estimate of $(8.5\pm3.0)\times 10^{11} M_{\odot}$ for the
total mass of M101, which is consistent with a ratio $16\pm6$ for the total
mass to the stellar mass of the galaxy.
\end{abstract}
\maketitle

\section{Introduction.}

M101=NGC~5457 is one of the closest massive galaxies of the late Scd type in
which there is essentially no spheroidal stellar component (bulge). These
galaxies usually lie in regions of low cosmic density. The amount of dark
matter in Sc and Sd galaxies per unit stellar mass is a factor of 2--4
smaller than in the spirals of earlier types (Sa,Sb) or elliptical galaxies
[1]. Dwarf spheroidal satellites with a low surface brightness (dSph),
lacking gas, and young stars are encountered comparatively rarely around 
Sc and Sd galaxies.

  Until
  recently, only one dSph- type galaxy, UGC~8882 [2], was known within
the vicinity of M101. A group of amateur astronomical photographers (Tief
Belichtete Galaxies) completed a survey of the surroundings of M101 on small
telescopes with very long exposure times and discovered 10 candidate dSph
satellites of M101 [3]. Some of these objects were rediscovered independently
by another team using a system of small high-throughput telescopes, the
''Dragonfly Telephoto Array'' [4]. A subsequent survey of the virial zone of
M101 at the 3.6-m CFHT telescope [5] resulted in the detection of another 38 
new candidate spheroidal satellites of M101 with typical apparent magnitudes
$B\simeq(19-21)^m$ and angular diameters $\sim(10-20)^{\prime\prime}$. Another 5 candidate dSph
satellites of M101 have been found [6] in a survey of a wider region of the
sky.

  Confirmation of the membership of spheroidal dwarfs in the M101 group by
measuring the radial velocity is complicated observational task because of
low surface brightness of these objects and the absence of gas and visible
structural features in them. It turned out to be possible, however, to
measure the distances of dSph galaxies on the Hubble Space Telescope (HST)
based on the luminosity of the tip of the red giant branch (TRGB). Images
of seven of the ten dwarf galaxies discovered in Refs. 3 and 4 were obtained
with the ACS camera on the HST in the GO 13682 program (PI P. van Dokkum).
An analysis of the data showed that three of the dwarf objects were real
satellites of M101 [7], while the remaining four were diffuse dwarf systems 
not resolved into stars [8].

  Carlsten et al. [9] have recently used data from the CFHT survey of the
surroundings of M101 to determine the distances of the previously found
spheroidal dwarfs by measuring the fluctuations in their surface brightness.
They concluded that most of these objects (29 out of 43) are distant galaxies
with no relationship to the M101 group. Besides the previously confirmed
three dSph satellites, DF1, DF2, and DF3, and the brighter galaxy UGC~8882,
only two of the spheroidal dwarfs, DwA and Dw9, were certified as new likely
members of the M101 group.

 \section{The distances of two new satellites of M101.}

  The archives of the HST contain images of 19 dwarf galaxies in the
neighborhood of M101 obtained with the HST ACS camera in F814W and F606W
filters as part of the program GO 14796 (PI D. Crnojevich) with exposures,
respectively, of 900 and 1200 s. We processed these images, taken from HST 
archive, by the standard procedure using the program package DOLPHOT [10].
\begin{figure}
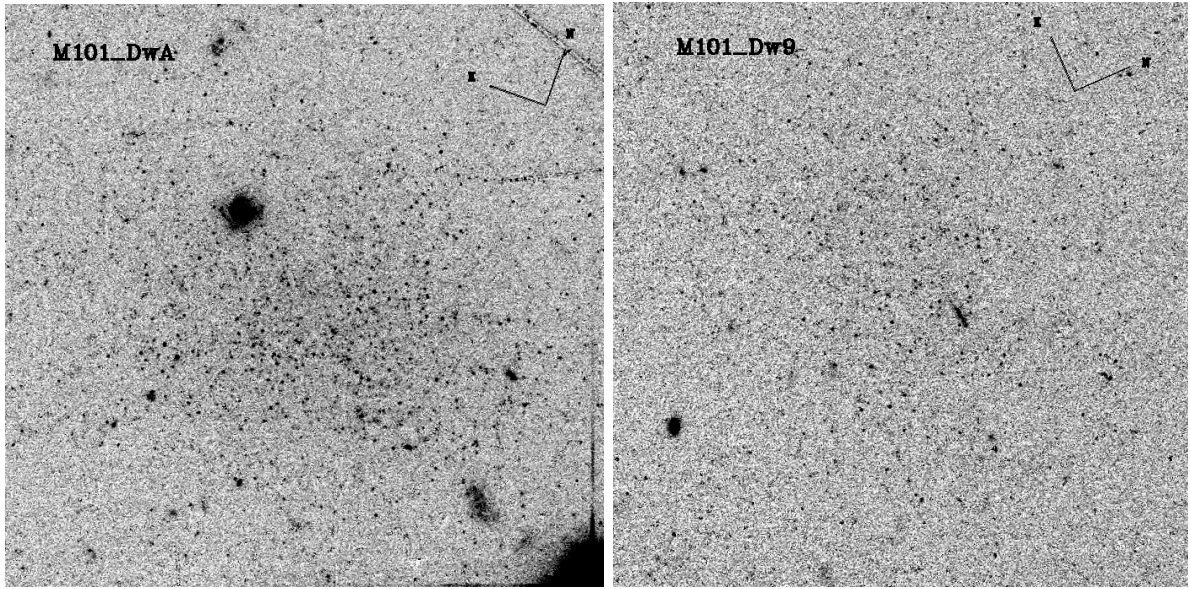
 %fig1
\setcaptionmargin{5mm} \onelinecaptionsfalse \captionstyle{normal}
\includegraphics[scale=0.5]{fig1a.ps}
\includegraphics[scale=0.48]{fig1b.ps}
\caption{Images of dwarf spheroidal galaxies in an F606W filter obtained with the ACS
camera of the Hubble Space Telescope. The size of the images is $35^{\prime\prime}\times35^{\prime\prime} $
and the north and east directions are indicated by arrows.}
\end{figure}

\begin{figure} %fig1
\setcaptionmargin{5mm} \onelinecaptionsfalse \captionstyle{normal}
\includegraphics[scale=0.5]{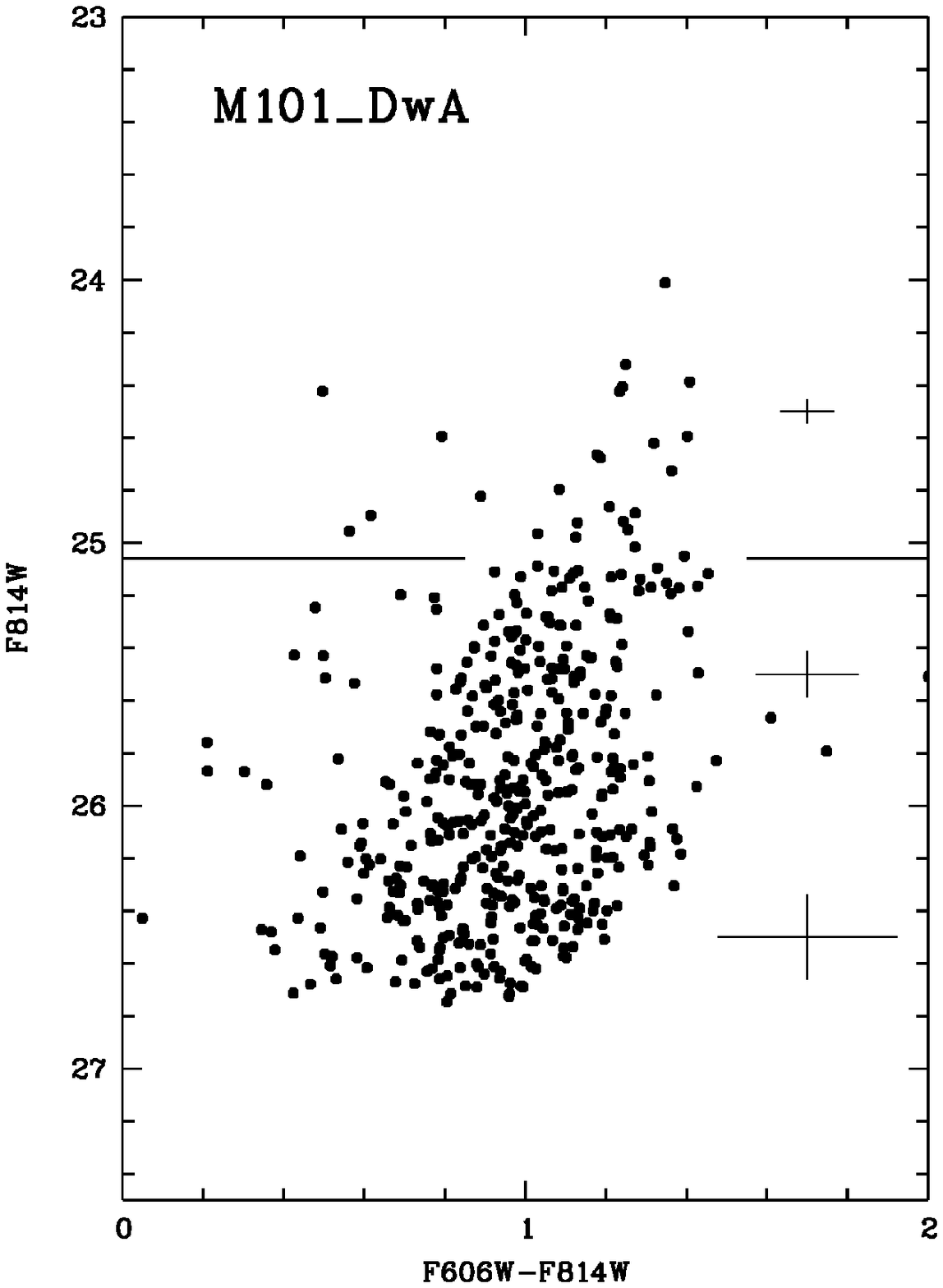}
\includegraphics[scale=0.5]{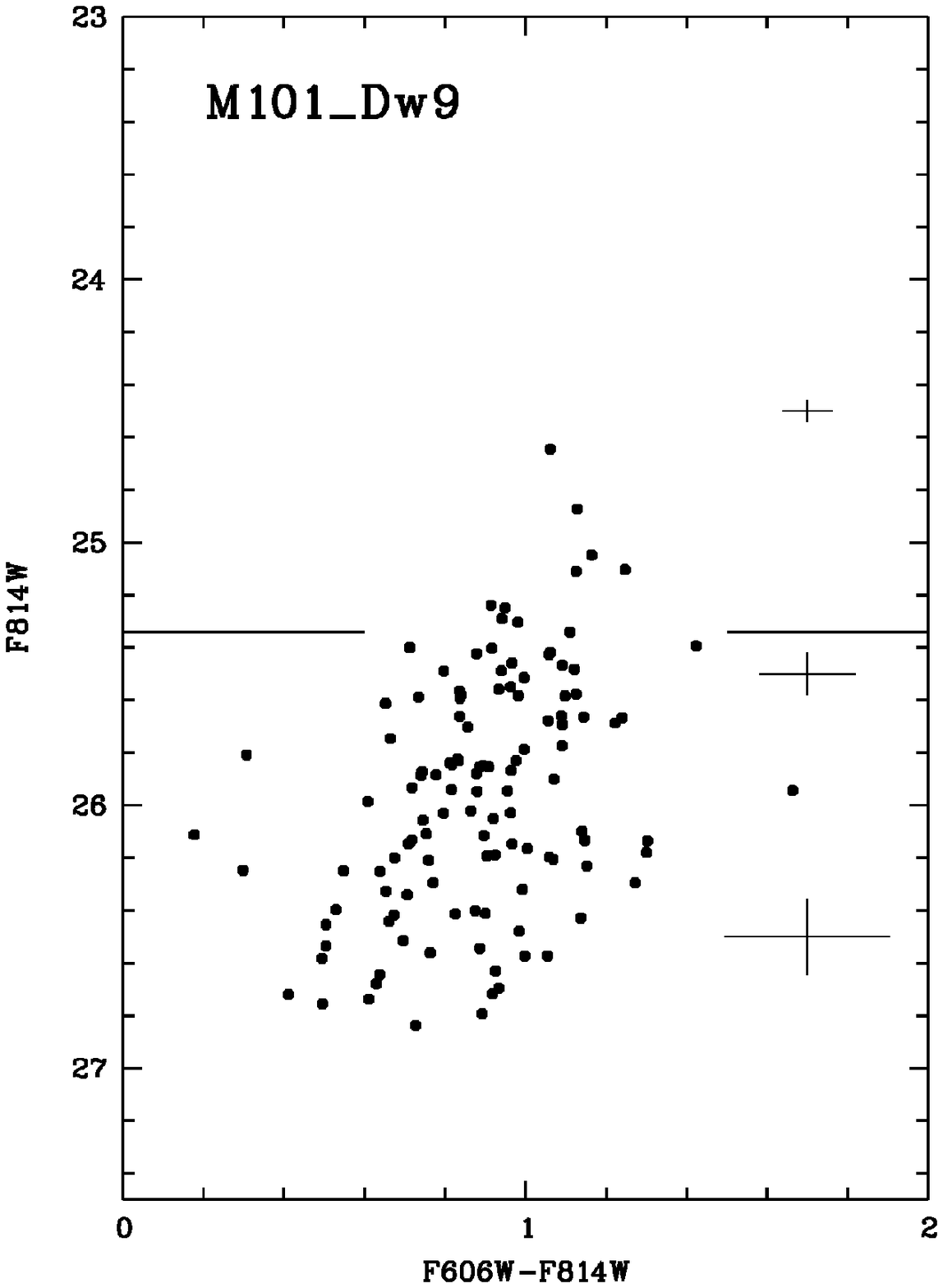}
\caption{Color-magnitude diagrams for the galaxies M101-DwA and M101-Dw9. The position of the TRGB
is indicated by a horizontal line. The crosses on the right show the photometric errors in the measurements.}
\end{figure}
During processing, the parameters recommended for elimination of nonstellar
and blended objects were used. In most cases the galaxies did not manifest
signs of resolution into stars. Only two of the galaxies, M101-DwA and M101-Dw9,
show up as distinctly resolved into stars, including a red giant branch, which 
can be used to determine the distances of these galaxies. Images of the two
galaxies are shown in Fig.1, where the two frames have an area of $35^{\prime\prime}\times 35^{\prime\prime}$.

  The tip of the red giant branch was found by a maximum likelihood principle
using modeling and subsequent recognition of artifical stars [11]. This
approach made it possible to estimate the photometric errors and the completeness
of the color-magnitude diagram (CMD) in fields with overlapping images of the
stars. The absolute magnitude of the TRGB was calibrated in accordance with
Ref. 12 taking the Galactic extinction from Ref.13 into account. CM diagrams
for both objects are shown in Fig. 2. The horizontal line indicates the
position of the TRGB and the crosses on the right are the photometric errors
determined from the artifical stars.

  These values of the TRGB, their errors, the magnitude of the assumed
extinction, and the distance moduli of the two dwarf galaxies are listed
in Table 1.
\begin{table}
 \caption{Estimated distances for the two dwarf galaxies.}
\begin{tabular}{cccccc}\hline

 Galaxy  &   TRGB   &   $A_I$    &  $(m-M)_0$      &  $D$,Mpc\\
\hline
 M101-DwA &   25.04  &  0.01  &   29.11$\pm$0.13 &  6.64\\
 M101-Dw9 &   25.36  &  0.01   &  29.44$\pm$0.13 &  7.71\\ \hline
 \end{tabular}
 \end{table}

  \section{The group of galaxies around M101.}

  At present, the group of galaxies associated with M101 has 15 members.
Data on them are given in Table 2. The columns of the table list: (1) the
galaxies; (2) their equatorial coordinates; (3) distances in Mpc; (4) the
method used to determine the distance: ''TRGB'' from the tip of the red giant
branch, ''sbf'' from surface brightness fluctuations, ''TF'' from the Tully-
Fisher relation [14], and ''mem'' from assumed membership in the M101 group;
(5) radial velocity relative to the centroid of the Local group; (6) 
logarithm of the stellar mass from the database http://www.sao.ru/lv/lvgdb;
(7) morphological type; (8,9) projected separation of the satellite in
degrees and in kpc; (10) references for data on the distance. The lower part
of the table includes 4 neighboring galaxies on the sky which are not
physical members of the group in terms of the distances.
\begin{table}
\caption{Major parameters of galaxies in the M101 group.}
\begin{tabular}{lcllcclcrl} \hline

 Galaxy     &  RA (2000.0) DEC&    D &  meth&  $V_{LG}$  &  $\log M^*$ &T & $r_p$&    $R_p$ &  Ref.   \\
            &                 &    Mpc &    &     km/s   & $M_{\odot}$    &      &  deg  &  kpc  &        \\
\hline
 dw1343+58  &134307.0+581340  & 6.95 &mem   &365    &7.65 & BCD  & 4.99 & 605   &       \\
 Holm IV    &135445.1+535417  & 6.93 &TRGB  &272    &8.60 & Sm   & 1.34 & 163  &[15]   \\
 GBT1355+54 &135450.6+543850  & 6.95 &mem   &345    &7.10 & HIcl & 1.25 & 152  &[16]   \\
 M101-Dw9   &135544.6+550845  & 7.71 &TRGB  & ---     &6.17 & Sph  & 1.37 & 166   &    \\
 UGC8882    &135714.6+540603  & 7.8  &sbf   &482    &7.64 & Sph  & 0.91 & 110  &[9]    \\
 M101-df3   &140305.7+533656  & 6.52 &TRGB  & ---     &7.27 & Sph  & 0.80 &  97  &[7]    \\
 M 101      &140312.8+542102  & 6.95 &TRGB  &378   1&0.79 & Scd  & 0.0  &   0  &[17]    \\
 M101-df1   &140345.0+535640  & 6.37 &TRGB  & ---     &6.17 & dTr  & 0.44 &  53  &[7]   \\
 PGC2448110 &140457.8+534129  & 6.95 &mem   &392    &6.91 & BCD  & 0.76 &  92   &       \\
 NGC5474    &140502.1+533947  & 6.82 &TRGB  &424    &9.20 & Sm   & 0.79 &  96  &[15]    \\
 NGC5477    &140533.1+542739  & 6.77 &TRGB  &451    &8.26 & Im   & 0.36 &  44  &[15]     \\
 M101-DwA   &140650.0+534429  & 6.64 &TRGB  & ---     &6.97 & Sph  & 0.84 & 102   &     \\
 M101-df2   &140837.5+541931  & 6.87 &TRGB  & ---     &6.74 & Sph  & 0.79 &  96  &[7]   \\
 NGC5585    &141948.3+564349  & 7.76 &TF    &457    &9.16 & Sd   & 3.54 & 429  &[17]    \\
 DDO194     &143524.6+571524  & 6.30 &TRGB  &381    &8.13 & Sm   & 5.33 & 647  &[15]    \\
\hline
 NGC5204    &132936.4+582504  & 4.59 &TRGB  &339    &8.84 & Sm   & 6.33 &  ---   &1[7]  \\
 UGC8508    &133044.4+545436  & 2.58 &TRGB  &181    &7.53 & dIr  & 4.73 &  ---   &[15] \\
 NGC5238    &133442.7+513650  & 4.21 &TRGB  &342    &7.99 & dIr  & 5.19 &  ---   &[5] \\
 KKH87      &141509.4+570515  & 9.40 &TRGB  &473    &7.76 & dIr  & 3.44 &  ---   &[15] \\
\hline 
\end{tabular}
\end{table}

These date   imply that the average projected separation for the 14 satellites
of M101 is 204 kpc, which is typical of the virial radius of the halo of
spiral galaxies of the type in the Milky Way and M31. The ''zero velocity 
radius'', which separates the group from the general cosmological expansion
is equal to $\sim1$ Mpc for galaxies of this virial mass [18]. Thus, even the
most distant dwarfs, DDO~194, NGC~5585, and dw1343+58, can be regarded as
physical members of the M101 group. The radial velocities were measured only
for the 9 brightest satellites of M101. The mean square difference in the
radial velocities is small for them and equals 64 km/s.

  Even the brightest satellites have a stellar mass an order of magnitude
and a half lower than the mass of the M101 itself. This suggests treating
them as a system of test particles moving around the central massive galaxy.
Assuming a random orientation of the orbits of the satellites with an average
eccentricity  e = 0.7, the average orbital estimate for the mass is given by

        $$ \langle M_{orb}\rangle=(16/\pi G)\langle\Delta V^2R_p\rangle,$$
where $G$ is the gravitational constant and $\Delta V$ is the radial velocity of the
satellite relative to M101 [19]. This Keplerian approach gives an average 
value for the mass of the halo of M101 of $(8.5\pm3.0) 10^{11} M_{\odot}$. For a stellar
mass of M101 equal to $5.3\times10^{10} M_{\odot}$, the ratio of the total mass of the halo
to the mass of the stellar component, $ \langle M_{orb}\rangle/M^* = 16\pm6$ turns out to be a
factor of two smaller than the typical value of ~$\sim32$ for M31, M81 and other
nearby spiral galaxies with developed bulges [19].

  \section{Groups of galaxies in the far background.}

  The abundance of dwarf spheroidal galaxies in the vicinity of M101 that are
not resolved into stars in the HST observations, requires an explanation. The
distribution of galaxies on their radial velocities in this sky region has a
substantial excess in the number of galaxies with radial velocities of
$\sim2000$ km/s. A catalog of groups of galaxies in the nearby universe [20]
includes $\sim350$ groups over the entire sky with average radial velocities
$V_{LG} < 3500$ km/s. Of these, there are 11 within a radius of 7 degrees around
M101. For all these groups the average radial velocities lie within an interval
of [1900 -- 2400] km/s. These groups form a diffuse structure of cosmic ''pancake''
type with a high probability. The two most populated groups around the galaxies
NGC~5422 and NGC~5473/85 lie so close in the sky to M101 that their zero 
velocity spheres encompass the galaxy M101 in projection. Table 3 lists the
following data on these groups: equatorial coordinates of the group center,
number of galaxies with the measured radial velocities, average radial velocity,
dispersion of of the radial velocities, distance in Mpc, virial radius,
integrated stellar and virial mass. The resulting ratios of the virial-to-stellar
masses  $M_v/M^* = 78$ and 33 are typical for groups with these morphological
populations.
\begin{table}
\caption{Parameters of the group of galxies behind M101.}
\begin{tabular}{cccrcccccc} \hline
 Group    &   J2000.0   & $N_v$ &$V_{LG}$&$\sigma_v$& D &$ R_v$ &$\log M^*$& $\log M_v$&  $M_v/M^*$\\
           &                 &  &km/s& km/s & Mpc  & kpc  & $M_{\odot}$ &  $M_{\odot}$&\\
\hline
NGC5422   & 140042.0+550952  &  12 & 1935  & 121  & 29.8  &  280 & 11.17 & 13.06 & 78\\
NGC5485   & 140711.3+550006  &  18 & 2162  & 94   & 27.9  & 277 & 11.20 & 12.72 & 33\\
\hline
\end{tabular}
\end{table}

\begin{figure} %fig1
\setcaptionmargin{5mm} \onelinecaptionsfalse \captionstyle{normal}
\includegraphics[scale=1.0]{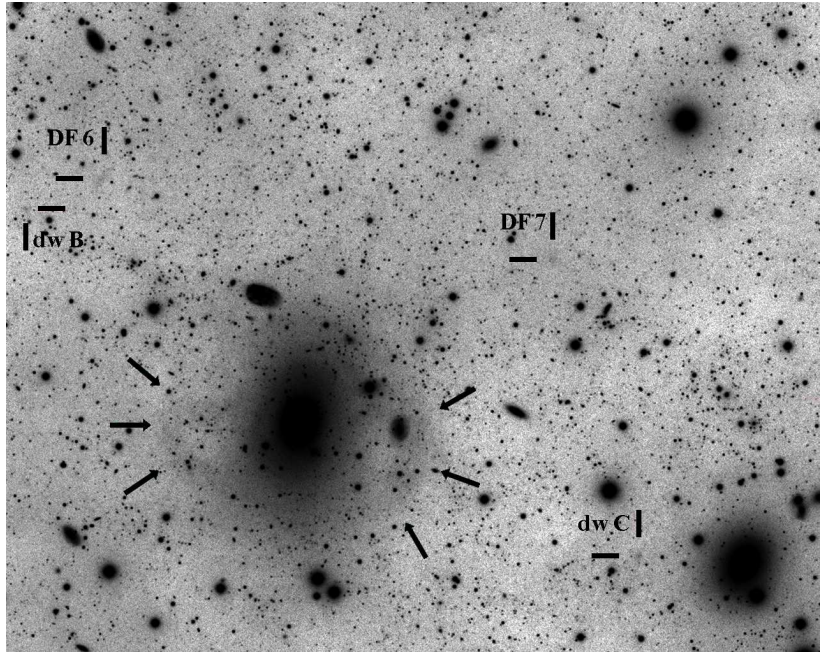}
\caption{The pair of S0 galaxies NGC~5473 (lower right) and NGC~5485 (near the center)
in the far group at a distance of 28 Mpc. The arrows indicate for 4 dwarf galaxies of this group, as  well as a system of stellar streams at the edge of NGC~5485. The frame size is $39^{\prime}\times 31^{\prime}$ and north is to the top, east to the left.}
\end{figure}
  Figure 3 is a fragment of the image of the central region of the group
NGC~5473/85 obtained by F.Neyer on the 15-cm amateur telescope with an
exposure time of 40 hours. Four dwarf galaxies of very low surface brightness
that were unresolved into stars in the HST images are indicated on this
figure. A system of faint peripheral streams indicated by arrows can be seen 
around the massive S0 galaxy NGC~5485. As members of the group at a distance
of $D = 27.9$ Mpc, the low-surface brightness dwarf objects have luminosities
and linear diameters that are typical of dSph galaxies in the more studied
nearby groups.

  As a whole, the population of dwarf spheroidal galaxies has a large variety
of luminosities and linear sizes [4,8]. The extreme manifestations of this 
class of objects are ultradiffuse and ultracompact dwarfs whose surface
brightnesses differ by hundreds of times. The large dispersion in the integrated
parameters of dSph galaxies makes it extremely difficult to estimate their
distances if they are far away and not resolved into separate stars.

  This work was supported by RFFI grant No. 18--02--00005. Data from the archive
of the Hubble Space Telescope have been used in this paper.

\bigskip

{\bf References}

\bigskip

 1.  Karachentsev I.D., Karachentseva V.E., 2019, MNRAS, 486, 3697

 2.  Rekola R., Jerjen H., Flynn C., 2005, A\&A, 437, 823

 3.  Karachentsev I.D., Riepe P., Zilch T., et al. 2015, Astrophysical Bulletin, 70, 379

 4.  Merritt A., van Dokkum P., \& Abraham R. 2014, ApJL, 787, L37

 5.  Bennet P., Sand D. J., Crnojevich D., et al. 2017, ApJ, 850, 109

 6.  Mueller O., Scalera R., Binggeli B., \& Jerjen H. 2017, A\&A, 602, A119 

 7.  Danieli S., van Dokkum P., Merritt A., et al. 2017, ApJ, 837, 136 

 8.  Merritt A., van Dokkum P., Danieli S., et al. 2016, ApJ, 833, 168

 9. Carlsten S.G., Beaton R.L., Greco J.P., Greene J.E., 2019, arXiv:1901.07578

 10. Dolphin, A. 2000, PASP, 112, 1383

 11. Makarov, D.I, Makarova, L., Rizzi, L. et al. 2006, AJ, 132, 2729

 12. Rizzi L., Tully R.B., Makarov D.I. et al. 2007, ApJ, 661, 813

 13. Schlafly, E.F., Finkbeiner, D.P. 2011, ApJ, 737, 103

 14. Tully R.B., Fisher J.R., 1977, A\&A, 54, 661

 15. Tikhonov N. A., Lebedev V. S., \& Galazutdinova O. A., 2015, Astronomy Letters, 41, 239

 16. Mihos J.C., Keating K., Holley-Bockelmann K. et al. 2012, ApJ, 761, 186

 17. Tully R.B., Courtois H.M., Sorce J.G., 2016, AJ, 152, 50  

 18. Kashibadze O.G., Karachentsev I.D., 2018, A\&A, 609A, 11 

 19. Karachentsev I.D., Kudrya Y.N., 2014, AJ, 148, 50 

 20. Makarov D.I., Karachentsev I.D., 2011, MNRAS, 412, 2498
\end{document}